\documentclass{ws-p8-50x6-00}
\def\lsim{\raise0.3ex\hbox{$<$\kern-0.75em\raise-1.1ex\hbox{$\sim$}}}
\def\gsim{\raise0.3ex\hbox{$>$\kern-0.75em\raise-1.1ex\hbox{$\sim$}}}

\newcommand{\bqa}{\begin{eqnarray}}
\newcommand{\eqa}{\end{eqnarray}}
\newcommand{\bc}{\begin{center}}
\newcommand{\ec}{\end{center}}

\begin{document}

\title{QCD THERMODYNAMICS \\
WITH 2 and 3 QUARK FLAVORS$^*$ 
}

\author{F. Karsch, E. Laermann, A. Peikert, Ch. Schmidt, S. Stickan}

\address{Fakult\"at f\"ur Physik, Universit\"at Bielefeld\\
33615 Bielefeld, Germany}

\maketitle

\abstracts{We discuss the flavor dependence of the pressure and 
critical temperature calculated in QCD with 2, 2+1 and 3 flavors 
using improved gauge and staggered fermion actions on lattices with
temporal extent $N_\tau=4$. For $T\; \gsim \; 2 T_c$ we find that bulk 
thermodynamics of QCD with 2 light and a heavier strange quark is well 
described by 3-flavor QCD while the transition temperature is closer
to that of 2-flavor QCD. Furthermore, we present evidence that the
chiral critical point of 3-flavor QCD, {\it i.e.} the second order
endpoint of the line of first order chiral phase transitions, belongs 
to the universality class of the 3d Ising model. }

\section{Introduction}
The existence of a finite temperature phase transition in strongly
interacting matter is one of the most exciting non-perturbative
features of QCD. Determining the equation of state and the transition 
temperature 
is one of the basic goals in
lattice studies of finite-T QCD. Studies of the transition which have
been performed during the last years have shown that the details of
the transition strongly depend on the number of quark flavors ($n_f$) as 
well as the value of e.g. the pseudo-scalar meson mass ($m_{\rm PS}$) 
which is controlled through variation of the bare quark masses ($m_q$). 
Furthermore, it became evident that the moderate 
values of the lattice spacing ($a \sim 0.25$fm) used in finite-T 
calculations with dynamical fermions lead to sizeable cut-off effects.
Different discretization schemes used in the fermion sector, e.g.
the standard staggered and Wilson fermion actions, lead to significantly 
different results for $T_c$ \cite{Kar99}. Calculations with improved 
actions thus seem to be mandatory to perform quantitative studies of 
the QCD equation of state at high temperature and to determine
accurately the value of $T_c$ and its dependence on $n_f$ and $m_{\rm PS}$.
We will report here on results for $n_f =2$ and 3 obtained with 
a Symanzik improved gauge action and an improved staggered fermion action,
the p4-action with fat 1-link terms\cite{Kar00}. In addition we present
results from a calculation with two light and a heavier strange quark.
Details on the improved action as well as
the algorithm used in our simulations are given in Ref.~2.

\vspace*{0.1cm}
\hrule

\vspace*{0.1cm}
\noindent
$^*$ Presented at the conference on {\it Strong and Electroweak Matter},
SEWM 2000, Marseille, June 13-17th, 2000.
Work supported by the TMR-Network grant ERBFMRX-CT-970122
and the DFG grant Ka 1198/4-1.

\vfill 
\eject

\section{Flavor dependence of the QCD equation of state}

\begin{figure}
\begin{center}
\epsfig{file=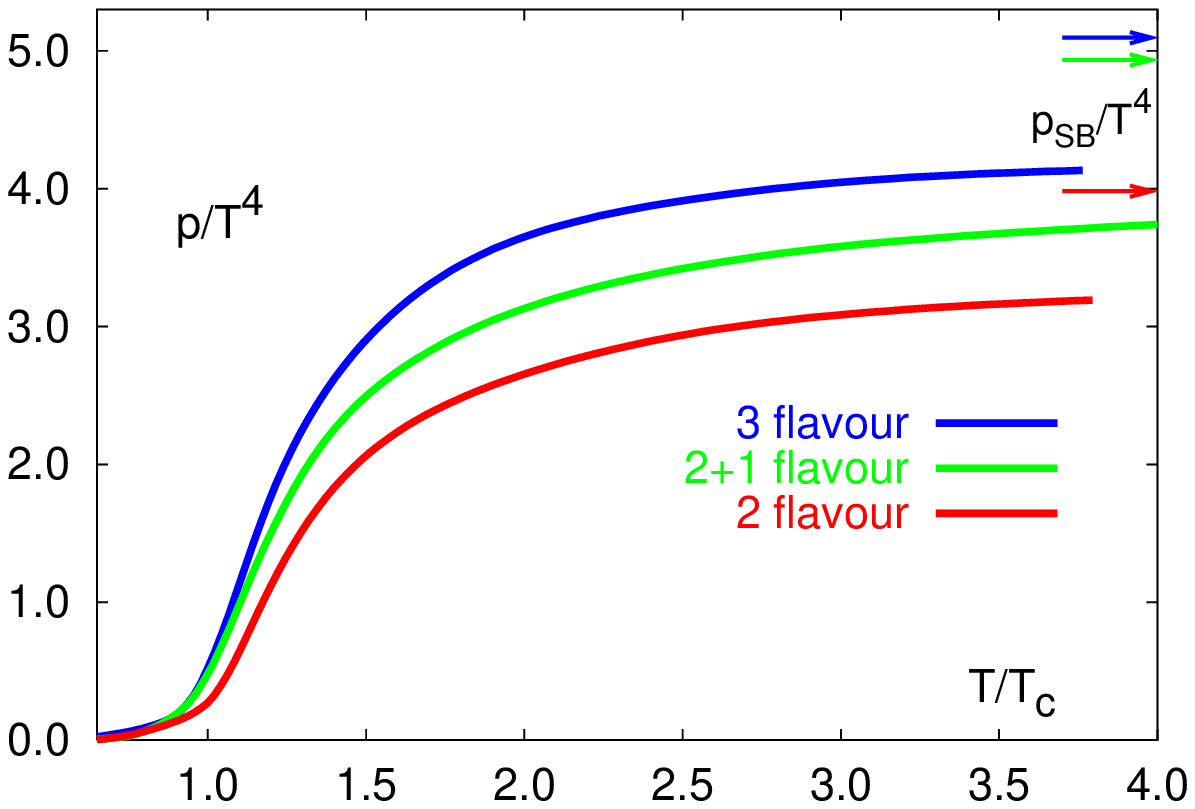,width=58mm}
\epsfig{file=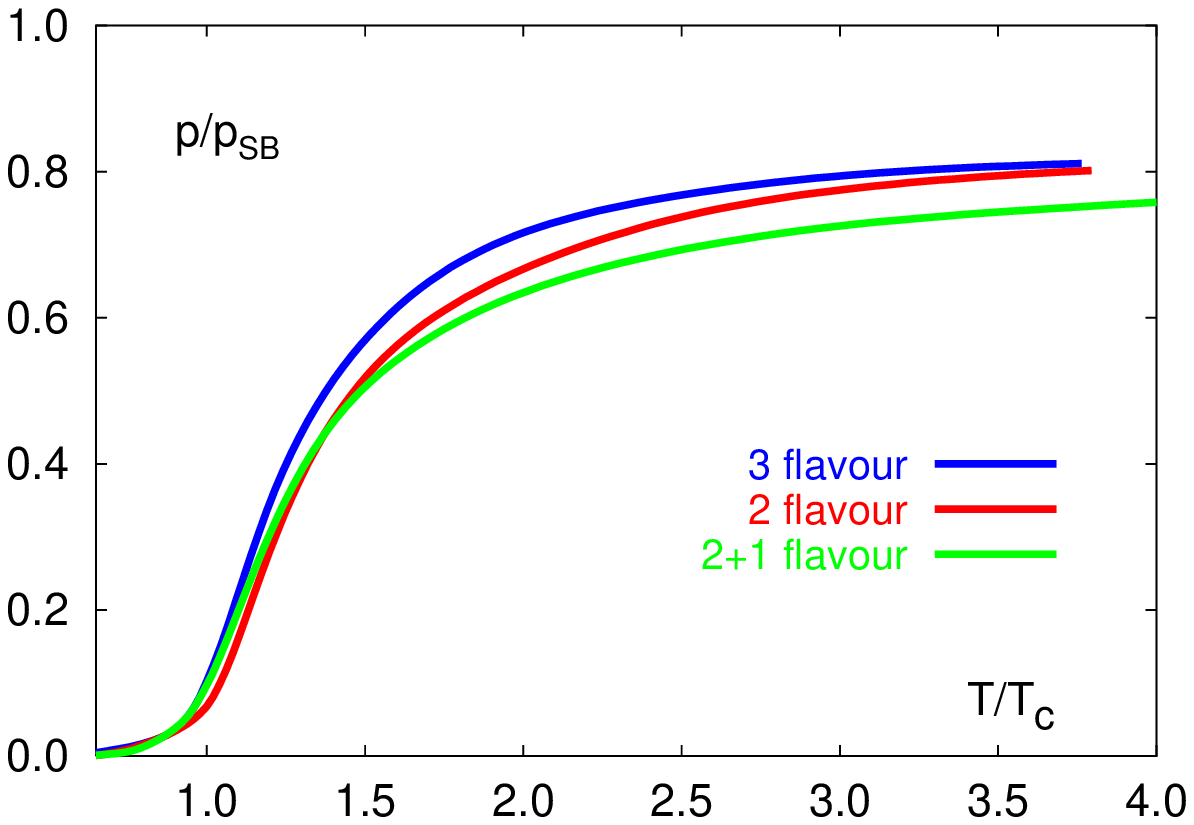,width=58mm}
\end{center}
\vspace*{-0.3cm}
\caption{The pressure in units of $T^4$ (left) and $p_{\rm SB}$ (right)
for 2, 2+1 and 3-flavor QCD obtained from calculations with the p4 
action on lattices with temporal extent $N_\tau=4$. The light quark mass
used in all cases is $m_q/T=0.4$ and the heavier quark mass used in the
(2+1)-flavor case is $m_s/T=1$.}
\label{fig:pressure}
\vspace*{-0.3cm}
\end{figure}

In the high (infinite) temperature limit the QCD equation of state is 
expected to approach that of an ideal quark-gluon gas, {\it i.e.} bulk
thermodynamic observables like energy density and pressure will 
reflect the number of light degrees of freedom,
\begin{equation}
\epsilon_{\rm SB}/T^4 = 3p_{\rm SB}/T^4 = \biggl( 8 + {21\over 4} g_f \biggr) 
{\pi^2 \over 15}
\quad ,
\label{eossb}
\end{equation}
where $g_f=\sum_{i=u,d,..} g(m_i/T)$ with
\begin{equation}
g(m/T)= {360\over 7\pi^4} \int_{m/T}^{\infty}{\rm d}x x
\sqrt{x^2-(m/T)^2} \ln\bigl(1+{\rm e}^{-x}\bigr)\quad ,
\end{equation}
counts the effective number of
degrees of freedom of a massive Fermi gas.
For a gas of massless quarks one has, of course, $g_f=n_f$. The
effective number of degrees of freedom, $g(m/T)$, rapidly approaches
unity for quark masses smaller than $T$. For instance, one has
$g(1)=0.8275$ and $g(0.4)=0.9672$. These numbers correspond to the bare
quark mass values used in our simulations\cite{Kar00} on
lattices of size $16^3\times 4$.
Results obtained for the pressure are shown in Fig.~\ref{fig:pressure}.
In the high-T phase $p/T^4$
clearly shows the expected flavor dependence. We note 
that lattice calculations are performed at fixed $m_q/T$. Using instead a 
fixed physical quark mass value, e.g. $m_s \simeq T_c$,
would require to reduce $m_q/T$ in a simulation as the temperature
is increased. From Fig.~\ref{fig:pressure} we thus conclude that 
already at $T\simeq 2T_c$ the pressure calculated in (2+1)-flavor QCD
with a fixed strange quark mass will be similar to that of massless 
3-flavor QCD.

\section{Flavor and quark mass dependence of $\bf T_c$}

The transition temperature in QCD with dynamical fermions has been
found to be significantly smaller than in the pure gauge sector. This
is in accordance with intuitive pictures of the phase transition based
e.g. on the thermodynamics of bag or percolation models. 
With decreasing $m_q$ the hadrons become lighter and it becomes easier 
to build up a sufficiently high particle density that can trigger a 
phase transition. For the same reason such models also suggest that $T_c$ 
becomes smaller when $n_f$ and in turn the number of light
pseudo-scalar mesons increases. This qualitative picture is confirmed by 
the lattice results presented below.

\begin{figure}
\begin{center}
\epsfig{file=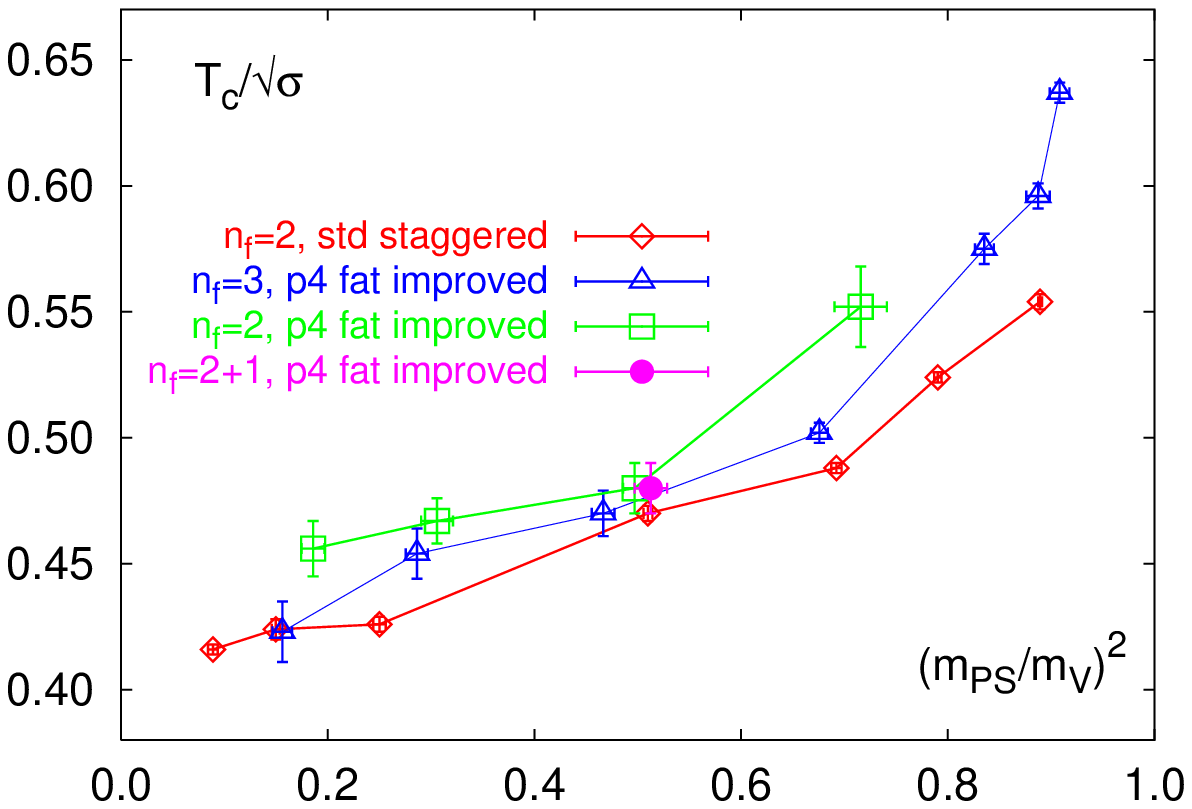,width=58mm}
\epsfig{file=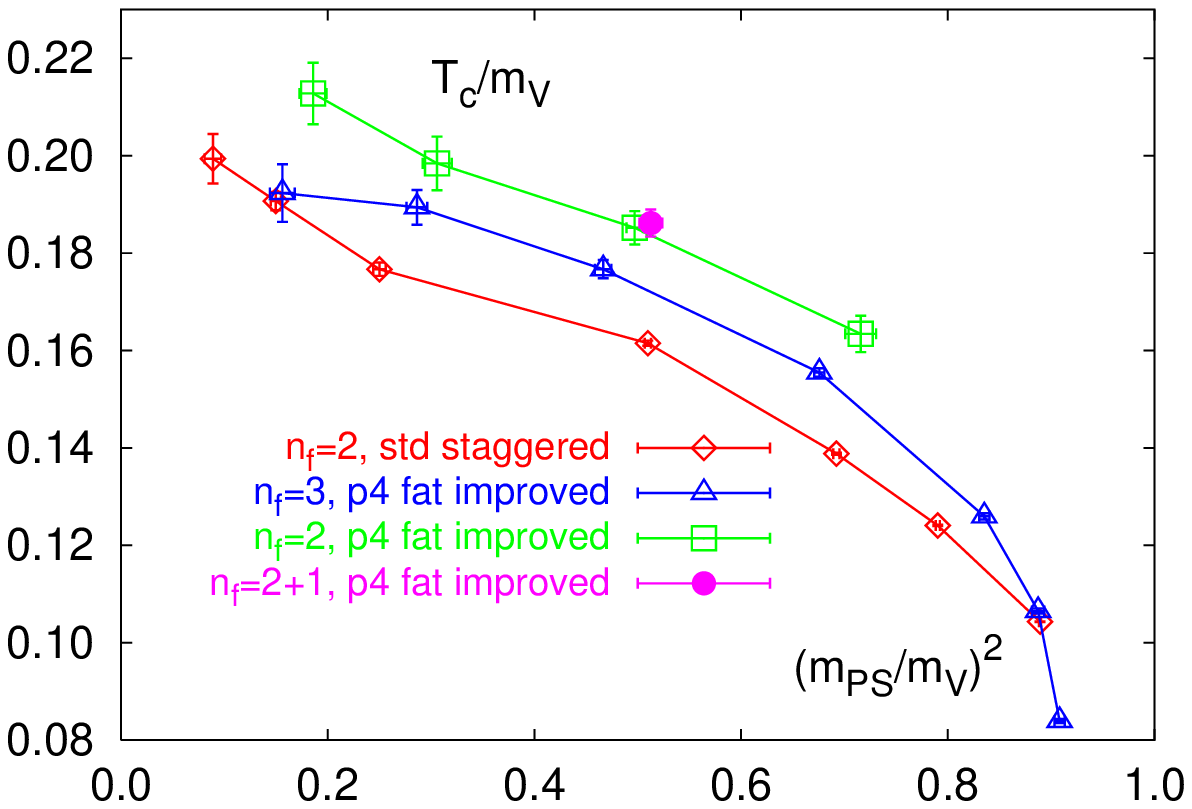, width=58mm}
\end{center}
\vspace*{-0.3cm}
\caption{The critical temperature in units of $\sqrt{\sigma}$  
(left) and the vector meson mass (right) versus 
$(m_{\rm PS}/m_{\rm V})^2$. Shown are results for 2, (2+1) and 3-flavor 
QCD obtained from calculations with the p4 action
on lattices with temporal extent $N_\tau=4$. For $n_f=2$ we also show
results obtained by using unimproved gauge and staggered fermion
actions.}
\label{fig:crit_temp}
\end{figure}

We have determined the pseudo-critical couplings, $\beta_c(m_q)$, 
for the transition to the
high temperature phase on lattices of size $16^3\times 4$. For 2 and 3-flavor 
QCD calculations have been performed in a wide range of quark masses, 
$0.025 \le m_q \le 1.0$. The smallest quark mass corresponds to 
a pseudo-scalar meson mass $m_{\rm PS} \simeq 350$~MeV. In order to set a 
scale for the transition temperature we  calculated the light
meson spectrum and the string tension\footnote{We note that the heavy quark
potential is no longer strictly confining in the presence of dynamical
quarks. The definition of the string tension is based on potentials
extracted from Wilson loops which have been found to lead to a linear 
rising potential at least up to distances $r\sim 2$fm.}
at $\beta_c(m_q)$ on lattices of size $16^4$. The resulting pseudo-critical
temperatures are shown in Fig.~\ref{fig:crit_temp}.

We note that $T_c/\sqrt{\sigma}$ and $T_c/m_V$ do show a consistent
flavor dependence of $T_c$. In a wide range of quark mass values, $T_c$ 
in 3-flavor QCD is about 10\% smaller than in 2-flavor QCD. 
The dependence on $m_q$ is, however, quite different when using the vector 
meson mass rather than the string tension to set the scale for $T_c$.  
While $T_c/\sqrt{\sigma}$ does show the expected rise with increasing
values of $m_q$ the contrary is the case for $T_c/m_V$. Of
course, this does not come as a surprise. The vector meson mass used in
Fig.~\ref{fig:crit_temp} to set the scale is itself strongly quark mass
dependent, $m_V = m_\rho +c\; m_q$. Its mass thus is significantly larger 
than the vector meson mass, $m_\rho$, in the chiral limit. 
We stress this well
known fact here because it makes evident that one has to be careful
when discussing the dependence of $T_c$ on parameters of the QCD 
Lagrangian, e.g. $n_f$ and $m_q$. One has 
to make sure that the observable used to set the scale for $T_c$
itself is not or at most only weakly dependent on the external parameters.
The fact, that the hadron spectrum as well as the string tension
calculated in quenched ($m_q\rightarrow \infty$) QCD\footnote{E. g.
the ratio $\sqrt{\sigma}/m_\rho$ calculated in the limit of vanishing
valence quark mass does show little dependence on the dynamical quark
mass used to generate gauge field configurations.} are in reasonable
agreement with experiment and phenomenology and show systematic deviations 
only on the 10\% level indicates that the corresponding partially quenched
observables are suitable also for setting the scale in the presence of 
dynamical quarks.

\section{$\bf T_c$ in the chiral limit}
The chiral phase transition in 3-flavor QCD is known to be first order
whereas it is most likely a continuous transition for $n_f=2$.
This also implies that the dependence of the (pseudo)-critical
temperature on the quark mass will differ in both cases. 
Asymptotically, {\it i.e.}
to leading order in the quark mass one expects to find, 

\begin{equation}
T_c (m_\pi) -T_c(0) \sim \cases{ m_\pi^{2/\beta\delta} &, $n_f=2$ \cr
 m_\pi^2 &, $n_f \ge 3$}\quad ,
\label{tcscaling}
\end{equation}
with $1/\beta\delta=0.55$ if the 2-flavor transition indeed belongs to the
universality class of 3d, O(4) symmetric spin models. Our estimates of 
$T_c$ in the chiral limit are based on the data shown in Fig.~2.
In the case of $n_f=3$ we have extrapolated $T_c/\sqrt{\sigma}$ and $T_c/m_V$ 
using an ansatz quadratic in $m_{\rm PS}/m_V$. In addition we have
extrapolated $m_V$ calculated at $\beta_c(m_q)$ for $m_q=0.025$ and 0.05
linearly in $m_q$ to the critical point in the chiral limit,
$\beta_c(0) = 3.258(4)$. The extrapolation to the chiral limit is 
less straightforward in the case of $n_f=2$. The data shown in Fig.~2 indicate
that the quark mass dependence for $n_f=2$ and 3 is quite similar. This
suggests that sub-leading corrections, quadratic in $m_{\rm PS}/ m_V$,
should be taken into account in addition to the leading behavior 
expected from universality. In our extrapolations for $n_f=2$ we thus
also add a term quadratic in $m_{\rm PS}/m_V$ to the leading term given
in Eq.~3.

\noindent
From these fits we estimate
\begin{equation}
{T_c \over m_\rho} = \cases{ 0.225 \pm 0.010 &, $n_f=2$ \cr
0.20 \pm 0.01 &, $n_f = 3$}\quad ,
\end{equation}
which corresponds to $T_c=(173 \pm 8)~{\rm MeV}$ and
$(154 \pm 8)~{\rm MeV}$ for $n_f=2$ and 3, respectively. 
For $n_f=2$ this agrees well with results obtained in a calculation with
improved Wilson fermions\cite{Ali00}. We stress, however, that
the errors given here as well as in Ref.~3 are statistical only. 
Systematic errors resulting
from remaining cut-off effects and from the ansatz used for extrapolating 
to the chiral limit are expected to be of similar magnitude. In order
to control these errors calculations on lattices with larger
temporal extent are still needed.

Physically most relevant is a determination of the transition temperature 
of
QCD with 2 light quarks ($m_{u,d}\simeq 0$) and a heavier (strange) 
quark with $m_s\simeq T_c$. The result from our calculation with 
$m_{u,d}/T=0.4$ and $m_s/T=1$ is also shown in Fig.~2.
Although this analysis of (2+1)-flavor QCD has not yet
been performed with a sufficiently light light
quark sector the result obtained for the pseudo-critical temperature
does suggest that also in the case of realistic light 
quark masses $T_c$ will be close to that of 2-flavor QCD. 

\section{Universality at the chiral critical point of 3-flavor QCD}
In the chiral limit of 3-flavor QCD the phase transition is first 
order\cite{old3}.
It thus will persist to be first order for $m_q > 0$ up to a
critical value of the quark mass, $m_c$. At this {\it chiral critical
point} the transition will be second order. It has been conjectured
that this critical point belongs to the universality class of the
3d Ising model\cite{Gav94}.  From a simulation with standard, {\it i.e.} 
unimproved, gauge and staggered fermion actions we find support for this
conjecture.

\begin{figure}[htb]
\vspace*{-1.7cm}
\begin{center}
\hspace*{-0.8cm}
\epsfig{file=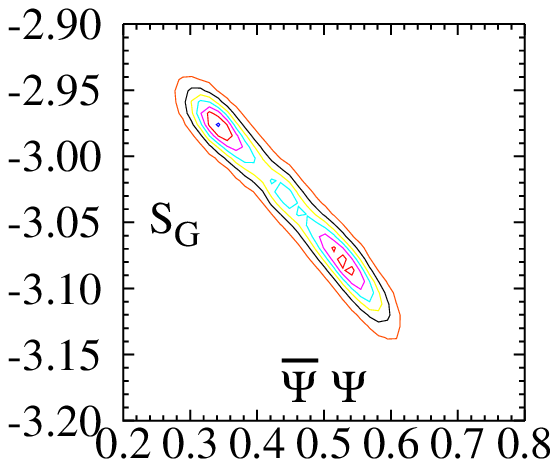,width=66mm}\hspace*{-2.7cm}
\epsfig{file=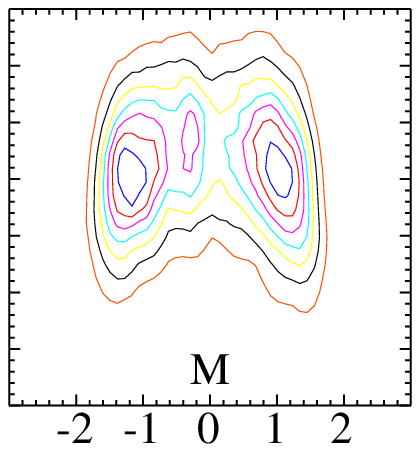,width=66mm}\hspace*{-3.5cm}
\epsfig{file=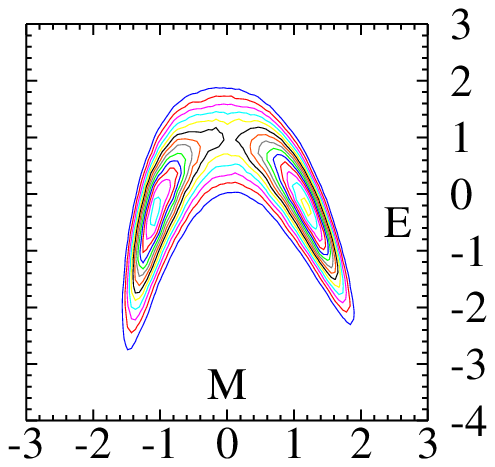,width=66mm}
\end{center}
\vspace*{-1.5cm}
\label{fig:histo_EM}
\caption{Contour plot for the joint probability distribution of $S_G$
and $\bar{\psi}\psi$ (left) as well as $E$ and
$M$ (middle) for 3-flavor QCD and the 3d, 3-state Potts model (right).
The QCD contour plots are based on calculations 
performed on a $16^3\times 4$ lattice at $\beta=5.1499$ with 
$m_q=0.035$. The mixing parameters have been 
fixed to $r=0.55$ and $s=0$. Parameters for the Potts model are given in
Ref.~7. 
}
\end{figure}

As discussed in the context of the electroweak transition the analysis
of the critical behavior at the $2^{\rm nd}$ order endpoint of a line of
$1^{\rm st}$ order phase transitions requires the correct identification of
energy-like and ordering-field (magnetic) directions\cite{higgs}. 
A general approach to determine the corresponding operators has been
discussed in the context of the 3d, 3-state Potts model\cite{Stickan}.
The ordering field operator at the chiral critical point
can be constructed from a linear combination of the gluonic action $S_G$
and the chiral condensate $\bar{\psi}\psi$,
\begin{equation}
E = S_G + r\; \bar{\psi}\psi \quad,\quad 
M = \bar{\psi}\psi + s\; S_G \quad.
\label{em}
\end{equation}
Here the mixing parameter $r$ is determined from the $m_q$-dependence
of the line of first order transitions, $r=({\rm d} \beta /{\rm d} m_q
)^{-1}_{\rm endpoint}$ and $s$ by demanding $\langle E\cdot M\rangle =
0$. 
In fact, unlike for energy-like observables, e.g. critical amplitudes
that involve the thermal exponent $y_t$, the universal properties of
observables related to $M$ do not depend on the correct choice of 
the mixing parameter $s$ as long as the magnetic exponent $y_h$
is larger than $y_t$. The joint probability distribution for $E$ and $M$
characterizes the symmetry at the critical endpoint and 
its universality class. Contour plots for the $E$-$M$
distributions at the critical endpoints of 3-flavor QCD and the 
3d, 3-state Potts model as well as the corresponding plot
for the $S_G$-$(\bar{\psi}\psi)$ distribution are shown in Fig.~3. 
This shows that also in the QCD case a proper definition of energy-like
and ordering-field operators is needed to reveal the symmetry
properties of the chiral critical point. The joint distributions of $E$
and $M$ suggest the universal structure of the
$E$-$M$ probability distribution of the 3-d Ising model\cite{higgs},
although it is apparent that an analysis of 3-flavor QCD on larger lattices 
is needed to clearly see the ``two wings''
characteristic for the 3d Ising distribution.


\vspace*{-0.3cm}
\begin{thebibliography}{9}

\vspace*{-0.2cm}
\bibitem{Kar99} F. Karsch, Nucl. Phys. B (Proc. Suppl.) 83-84 (2000) 14.
\bibitem{Kar00} F. Karsch, E. Laermann and A. Peikert,
Phys. Lett. B478 (2000) 447.
\bibitem{Ali00} A. Ali Khan et al. (CP-PACS Collaboration),
hep-lat/0008011.
\bibitem{old3} R.V. Gavai, J. Potvin and S. Sanielevici, 
Phys.~Rev.~Lett.~58~(1987)~2519. 
\bibitem{Gav94} S. Gavin, A. Gocksch and R.D. Pisarski,
Phys. Rev. D49 (1994) 3079.
\bibitem{higgs} 
J.L. Alonso et al., Nucl. Phys. B405 (1993) 574;
K. Rummukainen, et al., Nucl. Phys. B532 (1998) 283.
\bibitem{Stickan} F. Karsch and S. Stickan, Phys. Lett. B488 (2000) 319.
\end{thebibliography}
\end{document}